\documentclass[aps,prb,reprint]{revtex4-1}
\usepackage[dvipdf]{graphicx}

\usepackage{here}

\usepackage{bm} % bold math

\usepackage{epstopdf,float} % necessary for [H]

\newcommand{\Fig}[1]{Fig.~\ref{fig:#1}}

\newcommand{\SI}{the Supplementary Material}

\newcommand{\Ref}[1]{Ref.~\onlinecite{#1}}

\newcommand{\Tmax}{T_\mathrm{max}}
\newcommand{\Nrep}{N_\mathrm{rep}}

\newcommand{\Nc}{N_c}
\newcommand{\Lp}{L_\mathrm{p}}

\newcommand{\Euu}{E_\mathrm{uu}}
\newcommand{\Euv}{E_\mathrm{uv}}
\newcommand{\Evv}{E_\mathrm{vv}}

\begin{document}

%.........................................................
% @title

\title{Self-assembly of a model supramolecular polymer studied by\\
replica exchange with solute tempering}

\author{Hadi H. Arefi}
\author{Takeshi Yamamoto}
\email{yamamoto@kuchem.kyoto-u.ac.jp.}
\affiliation{Department of Chemistry, Graduate School of Science,
Kyoto University, Kyoto 606-8502, Japan}

%------------------------------------------------------------------------
% @abst

\begin{abstract}
Conventional molecular-dynamics (cMD) simulation has a well-known
limitation in accessible time and length scales, and thus various enhanced
sampling techniques have been proposed to alleviate the problem.
In this paper we explore the utility of replica exchange
with solute tempering (REST) (i.e., a variant of Hamiltonian replica
exchange methods) to simulate the self-assembly of a supramolecular
polymer in explicit solvent, and compare the performance with
temperature-based replica exchange MD (T-REMD) as well as cMD.
As a test system, we consider a relatively simple all-atom model of
supramolecular polymerization (namely, benzene-1,3,5-tricarboxamides
in methylcyclohexane solvent).
Our results show that both REST and T-REMD are able to predict highly
ordered polymer structures with helical H-bonding patterns,
in contrast to cMD which completely fails to obtain such a structure
for the present model. At the same time, we have also experienced
some technical challenge (i.e., aggregation-dispersion
transition and the resulting bottleneck
for replica traversal), which is illustrated numerically.
Since the computational cost of REST scales more moderately
than T-REMD, we expect that REST will be useful for studying
the self-assembly of larger systems in solution
with enhanced rearrangement of monomers.
\end{abstract}

\maketitle

%------------------------------------------------------------------------
% @Intro

\textbf{Introduction.}
Supramolecular polymerization, i.e., the self-assembly of monomers
into one-dimensional ordered structures via non-covalent interactions,
has been receiving increased attention for developing advanced
functional materials.
\cite{deGreef_suprapoly_Nat08,
Meijer_suprapoly_review_ChemRev09,
Meijer_suprapoly_review_Sci12,
Bala_suprapoly_review_CPC13,
Zhang_suprapoly_review_ChemRev15}
A variety of factors including monomer structures, environments,
and their interactions play crucial roles for
assembly structures and aggregation properties.
To obtain more insights into their relationship,
molecular-dynamics (MD) simulations have been carried out
for representative systems.\cite{Bala_suprapoly_review_CPC13}
However, a well-known drawback of direct MD simulations
(particularly based on all-atom models)
is that accessible time and spatial scales are often too limited
to study self-assembly processes.
Therefore, various approaches have been used to alleviate the difficulty,
including the use of coarse-grained models.\cite{Martini_review13}

%[ T-REMD and REST ]
In this paper we are interested in the utility of enhanced sampling
\cite{Schulten_review_BBA15,Bussi_review_Entropy14,Wales_review_WIRE15}
for studying supramolecular polymerization.
Temperature-based replica exchange MD (T-REMD)\cite{Hansmann_REMD,Okamoto_REMD}
is among the most popular approaches for this purpose.
In T-REMD, one evolves a set of replicas at different temperatures
and exchanges their coordinates periodically.
This facilitates the replica at the target temperature
to overcome energy barriers and
explore a wider conformational space.\cite{Garcia_REMD_review}
Despite its utility, the computational cost of T-REMD
increases with the total degrees of freedom,
making it expensive to study a large system in explicit solvent.
Replica exchange with solute tempering (REST)
\cite{Berne_REST1,Berne_REST_test,Berne_REST2}
deals with this problem by using a modified potential
energy function\cite{Berne_REST2,Terakawa_REST_impl,Bussi_REST_impl}
\begin{equation}
  E_m =
    \frac{ \beta_m }{ \beta_0 } \Euu
  + \sqrt{ \frac{ \beta_m }{ \beta_0 } } \Euv
  + \Evv
  ,
\end{equation}
where $\Euu$, $\Euv$, and $\Evv$ are
solute-solute, solute-solvent, and solvent-solvent interaction
energies, respectively, $\beta_m = 1/(k_B T_m)$ with $T_m$
the effective temperature of replica $m = 0,\ldots,\Nrep-1$,
and $\Nrep$ is the number of replicas.
Note that each replica is run at the target temperature
[i.e., $\exp(-\beta_0 E_m)$ is sampled], which makes the solute-solute
and solute-solvent interactions effectively weaker in higher replicas.
The corresponding acceptance formula is independent of
$\Evv$, and hence a much reduced number of replicas.\cite{Berne_REST2}
The REST has been applied recently to several biological systems with
considerable success,
\cite{Tiana_REST_Prot08,Walsh_REST_PCCP13,Wang_REST_JCTC13,
Garcia_REST_JCTC14,Jorgensen_REST_JCTC14,
chignolin_REST_JCTC15,Zhou_REST_BJ15,Klimov_REST_JCTC16}
but to the best of our knowledge it has not been applied
to the self-assembly of dispersed monomers in solution.

%[ Purpose of this paper + BTA intro ]
The purpose of this paper is therefore to explore the utility
of REST for studying supramolecular polymers
in explicit solvent. For this purpose, we consider the assembly
of benzene-1,3,5-tricarboxamides (BTAs)\cite{deGreef_BTA_review}
in methylcyclohexane (MCH) solvent as a test system.
The BTA consists of a benzene core with three amide groups,
which form three-fold H-bonds and provide a major driving force
for columnar assembly.
Depending on alkyl side chains, aggregates of diverse morphology
can be obtained (see \Ref{deGreef_BTA_review} for a recent review).
A variety of theoretical studies have also been performed for related systems.
\cite{Bala_BTA_AA,Bala_BTA_CG,Bala_BTA_PCCP17,Bala_BTA_ChemComm15,
Pavan_BTA_JACS16,Pavan_BTA_Nano17,Pavan_BTA_JPCL17,
deGreef_BTA_DFT_JPCB10}
In the following we consider a relatively small system
(10 BTA monomers dissolved in 515 MCH molecules) to facilitate
extensive comparison among different methods. 
Nevertheless, the present system poses a significant challenge
to statistical sampling because of the competition between
``proper'' H-bonds between stacked monomers and ``improper''
H-bonds via side-by-side attachment (see below).
More computational details and additional data are provided in \SI{}.

%------------------------------------------------------------------------
% @cmd

\begin{figure}[h]
  \centering
  \includegraphics[width=8.5cm, clip]
                  {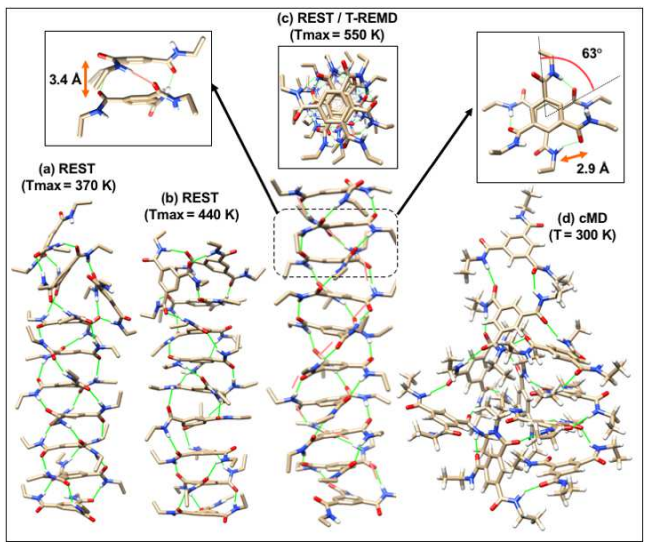}
\caption{\label{fig:struct}
Assembly structures obtained from cMD, T-REMD, and REST with
different values of $\Tmax$. While cMD gives an amorphous aggregate for the
present system [(d)], both REST and T-REMD with $\Tmax$ = 550 K
produce fully elongated BTA stacks with helical 2:1 H-bonding patterns
[(c)]. See \SI{} for more polymer structures obtained.
When $\Tmax$ is lower [(a) and (b)], a few monomers are misbound to the
terminal end of the polymer.
}
\end{figure}

\textbf{Conventional MD.}
To begin, we performed a conventional MD (cMD) calculation at 300 K.
The initial state of the cMD calculation was chosen
as a molecularly dispersed state.\cite{SI}
We find that the monomers rapidly form an amorphous aggregate
within $\sim$20 ns. A typical MD snapshot is shown \Fig{struct} (d).
As seen, the random aggregate involves many ``improper'' H-bonds
via side-by-side attachment of monomers.
We continued the cMD calculation up to 3000 ns, but
the system was not able to escape from the random state
and remained trapped in a meta-stable state (or a ``local minimum''
on the energy landscape). We repeated the same calculation
with a different random seed, but the system was again
trapped in a random state. The strong tendency toward amorphous
aggregates is due to the relatively short alkyl side chains
allowing side-by-side attachment and a rather high concentration
of monomers (as typical of all-atom assembly simulations).\cite{SI}
Fig.~S3 in \SI{} displays the evolution of the number
of clusters in the system (denoted as $\Nc$) at 300 K.
As seen, $\Nc$ starts from $\sim$10 (corresponding to a molecularly
dispersed state) and rapidly decays to 1 (a single aggregate),
in qualitative agreement with MD snapshots.
The system then retains $\Nc$ = 1 or 2 for the rest of the simulation.
%
%
%[ Assembly structures ] @fig @struct

%------------------------------------------------------------------------
% T-REMD
% @tremd

\textbf{T-REMD.}
%[ Minimal details ]
To avoid such kinetic trap, we next performed
the T-REMD calculation to obtain a reference result.
Here we set the lowest replica temperature ($T_0$) to 300 K
(i.e., the target temperature)
and the highest replica temperature
($\Tmax$) to 550 K.
The intermediate temperatures $\{ T_m \}$ were obtained
by using a web-based T-generator\cite{Tgen}
and requesting an average exchange rate of 20 \%.
This resulted in a total of 30 replicas spanning 300-550 K.
The potential energy distribution of each replica
shows sufficient overlap with each other,\cite{SI}
resulting in an actual exchange rate of 20-24 \%.
To make a fair comparison with cMD,
the T-REMD calculation was performed for 100 ns per replica
so that the total MD time becomes 3000 ns.

\begin{figure}[h]
  \centering
  \includegraphics[width=8.5cm, clip]
                  {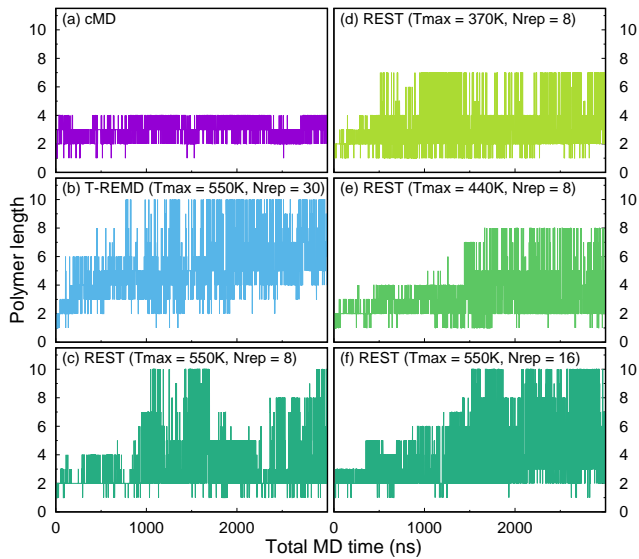}
\caption{\label{fig:polylen}
Evolution of $\Lp$ (i.e., the maximum length of neatly stacked
polymers) obtained from (a) cMD, (b) T-REMD,
(c-f) REST calculations with different choice of $\Tmax$ and $\Nrep$.
For more details of each calculation, see Table S1 in \SI.
}
\end{figure}

%[ Evolution of Lp ]
In the T-REMD method, the replica at the target temperature
samples a variety of configurations with the help of higher-temperature
replicas. Indeed, we find that the target replica
exhibits various assembled structures ranging from a partially
ordered aggregate to a fully elongated polymer.
To quantify the degree of order in the system,
we have calculated the length of the longest neatly stacked polymer,
denoted as $\Lp$ (see \SI{} for details).
With the present definition, $\Lp$ = 1 means that
no neatly stacked monomers exist in the system,
while $\Lp$ = 10 means that the monomers
form a fully elongated columnar structure.
\Fig{polylen} (b) displays the evolution of $\Lp$ for the replica at the
target temperature. It is seen that $\Lp$ takes on a wide range of values
(2--10), suggesting that the canonical ensemble at 300 K
is a mixture of partially ordered states
(i.e., not dominated by a single long polymer).

%[ Polymer structure / 2:1 H-bond pattern ]
\Fig{struct} (c) displays a typical assembly structure
corresponding to $\Lp$ = 10.
Interestingly, the monomers exhibit helical 2:1 H-bonding patterns.\cite{SI}
That is, one of the three amide hydrogens orient in one direction
of the polymer (e.g., ``up''), while the remaining amide hydrogens
are oriented in the other direction (``down'').
This H-bonding pattern agrees qualitatively
with a previous theoretical study
on closely related systems\cite{Bala_BTA_AA}
and suggests that it is energetically more stable than possible
3:0 H-bonding patterns (where all the amide hydrogens
are oriented in the same direction).

%[ Lp for cMD ]
\Fig{polylen} (a) displays the evolution of $\Lp$ for the cMD calculation
at 300 K. As seen, the value of $\Lp$ increases only up to 4,
which corresponds to a short polymer fragment
in an amorphous aggregate. The polymer does not grow further
because of insufficient rearrangement of monomers
within a given simulation time.

%------------------------------------------------------------------------
% @Rest

\textbf{REST.}
%[ Simulation details ]
We next applied the REST method to the present system.
For comparison, the lowest and highest replica temperatures
were chosen the same as the T-REMD calculation
(i.e., $T_0$ = 300 K and $\Tmax$ = 550 K).
The intermediate replica temperatures were determined by assuming
a simple geometric progression\cite{SI} and adjusting the number of replicas
($\Nrep$) to give an average exchange rate of $\sim$20 \%.
This procedure resulted in $\Nrep$ = 8 and actual
exchange rates of 15-22 \%. The effective potential
energy for each replica exhibits a relatively broad distribution.\cite{SI}
Thus, a smaller number of replicas suffice to span the same temperature range.
The simulation time was set to 375 ns (per replica)
to make the total MD time equal to the other calculations.

%[ Evolution of Lp / structure of polymer ]
\Fig{polylen} (c) displays the evolution of $\Lp$ obtained from
the REST calculation. As seen, the REST is also able to sample
a wide range of $\Lp$ and generate fully elongated
polymers with $\Lp$ = 10. The polymer structure thus obtained
is essentially the same as shown in \Fig{struct} (c).
The constituent monomers also exhibit the helical 2:1 H-bonding patterns.
Thus, we find that REST successfully reproduces the main feature of
supramolecular polymer as obtained from T-REMD.

%[ Statistical distributions ]
Some statistical data are shown in \SI{}.
The distribution of the number of H-bonds
and the radial distribution functions (RDFs) of benzene cores
show good agreement between T-REMD and REST. On the other hand,
the histogram of $\Lp$ exhibits more discrepancy between the two,
suggesting that longer sampling time is necessary for
better agreement. Indeed, the distribution of $\Lp$ is more
difficult to converge because the present definition of
$\Lp$ is rather tight\cite{SI} and thus it measures
the formation of highly ordered structures in the system.
Another factor that may affect the convergence rate
of REST is the envelope-like feature of $\Lp$ observed
in \Fig{polylen} (c), which will be discussed later.
%
%
%[ Polymer length ] @fig @polylen

%.........................................................
% @REST @lowerTmax

\textbf{REST with lower $\bm{T}_\mathrm{max}$.}
The performance of replica-exchange methods is affected by many
simulation parameters, and here we study the effect of
$\Tmax$ on assembly structures.
In replica-exchange simulations for biological systems,
it is typical to set $\Tmax$ to around 400--500 K.
We thus utilized somewhat lower values of $\Tmax$ (370 and 440 K).
For simplicity, the number of replicas was chosen the same
as the REST calculation above ($\Nrep$ = 8).
Because of the lower $\Tmax$, the effective energy distribution
shows even greater overlap between adjacent replicas,
resulting in a higher exchange rate on average
(66 and 46 \% for 370 and 440 K, respectively).\cite{SI}

%[ Tmax = 370 or 440 K : insufficient polymer growth ]
\Fig{polylen} (d) and (e) display the evolution of $\Lp$ obtained
with lower $\Tmax$. This figure shows
that $\Lp$ increases only up to 7 and 8
for $\Tmax$ = 370 and 440 K, respectively.
Typical polymer structures corresponding to $\Lp$ = 7 and 8
are displayed in \Fig{struct} (a) and (b).
As seen, two or three monomers bind erroneously
to one terminal end of the polymer in a non-stacked manner
and behave as a defect for further polymer growth.
Those monomers are not able to rearrange themselves
to a neatly stacked position, i.e., ``error correction''
does not occur via higher replicas.
Erroneous binding occurs preferentially
at the terminal end (rather than the side) of a polymer,
which may be due to the macrodipole
of the polymer\cite{deGreef_BTA_review,Bala_BTA_AA,Bala_BTA_ChemComm15}
and an enhanced electrostatic field at the terminus.
%
%[ Tmax = 550, Nrep = 16 : full polymer growth again ]
To make further comparison, we also performed the REST simulation
with $\Tmax$ = 550 K and $\Nrep$ = 16 to obtain a greater
exchange rate on average (58 \%). The evolution of $\Lp$
[\Fig{polylen} (f)] shows that full BTA stacks with $\Lp$ = 10
are obtained again, indicating that the result is reproducible
as long as $\Tmax$ is sufficiently high.

%[ Need for high Tmax ]
The above result suggests that the choice of $\Tmax$
is crucial for successful polymer elongation.
This observation is consistent with the idea that
$\Tmax$ should be high enough so that the system can overcome
energy barriers for the process of interest.\cite{Garcia_REMD_review}
In the present case the energy barrier arises from inter-monomer H-bonds,
and the system needs to break such H-bonds via high-temperature replicas.
However, the use of high $\Tmax$ also allows monomers to
dissociate into the bulk solvent. This has both pros and cons
for sampling efficiency: On one hand, the dissociation of monomers
allows for partial or total ``resetting'' of an aggregation process
at the target temperature, thus helping the system escape from
local minima. On the other hand, dissociated monomers necessarily
increase the entropy of the system, which is not favorable in REMD.
Thus, there is some dilemma as to the choice of $\Tmax$
for efficient simulation.
%
%
%[ Replica traversal (REST) ] @fig @traversal_rest
\begin{figure}[h]
  \centering
  \includegraphics[width=8.8cm, clip]
                  {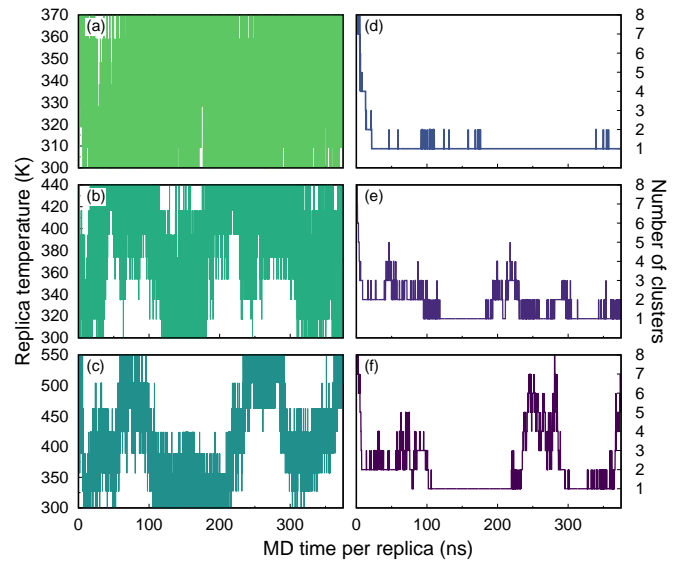}
\caption{\label{fig:traversal_rest}
Replica traversal in the temperature space for typical replicas
(left panels) and the corresponding evolution of $\Nc$ (right panels)
for the REST calculations.
Top, middle, and bottom panels correspond to
$\Tmax$ = 370, 440, and 550 K, respectively.
All the simulations were performed with $\Nrep$ = 8.
}
\end{figure}
%
%

%.........................................................
% @REST @lowerTmax
% @replica @traversal
%
\vspace{0.5cm}
\textbf{Aggregation-dispersion transition and
a bottleneck in replica temperature space.}
%
%[ Tmax = 370 K ]
The situation can be seen more clearly by examining the replica traversal
in temperature space.
\Fig{traversal_rest} (a) displays the evolution of temperature for
a typical replica with $\Tmax$ = 370 K. The round trip occurs very rapidly
between 300 and 370 K, so that the plotted curve almost fills the panel.
The value of $\Nc$ for the same replica [panel (d)]
starts from $\sim$10 (the initial dispersed state)
and decays rapidly to 1. The system then retains a single aggregate
throughout the simulation (except for rare occurrence of $\Nc$ = 2).

%[ Tmax = 440 and 550 K ]
When $\Tmax$ is raised to 440 K [panel (b)], the replica
starts to show a wave-like (or envelope-like) feature in temperature space.
The corresponding value of $\Nc$ indicates that the system
often decomposes into several clusters.
This trend becomes more evident for $\Tmax$ = 550 K [panel (c)].
Importantly, the oscillatory pattern of replica temperature
is closely related to the aggregation-dispersion transition
at around 400-450 K.
At low replica temperatures the system retains a single aggregate,
while at higher temperatures the replica makes a phase transition
to a dispersed state.
Once the monomers dissociate into the bulk, a certain time (20--50 ns) is
necessary for the recombination to occur.
This dissociation-recombination process limits the time scale of
``round trips'' of replicas in temperature space.
The above observation is somewhat analogous to the relation between
cooperative transition in proteins and the appearance
of a bottleneck in temperature space.\cite{Garcia_REMD_review}
That is, large temperature changes are slaved (or connected) to
conformational changes in the replicas.
In the present case, the dissociation and association
of monomers become a limiting factor for
the replica traversal between $T_0$ and $\Tmax$.

%[ T-REMD ]
T-REMD also exhibits a similar behavior by reflecting the aggregation-dispersion
transition at 400-450 K (see \SI{}). An interesting observation
is that the round trips occur somewhat faster for T-REMD than REST.
This is probably because the solvent is also heated in T-REMD,
which facilitates the diffusion of monomers at high temperatures.
This ``hot solvent'' effect contributes to the overall good efficiency
of T-REMD despite the greater number of replicas.
Nevertheless, we expect that REST will be advantageous
for larger systems because of more moderate scaling
of the number of replicas, which facilitates
replica simulation with given parallel
computational resources.\cite{REST_largesys}

%------------------------------------------------------------------------
% @concl

\textbf{Conclusions.}
In this paper we have explored the utility of REST
for supramolecular polymerization by using a relatively simple
model in explicit solvent. For the present system, cMD has produced
an amorphous aggregate and thus completely failed to predict assembly
structures. On the other hand, both REST and T-REMD
successfully produced a fully elongated polymer
with helical H-bonding patterns. To obtain such a structure,
it was necessary to raise the highest replica temperature ($\Tmax$)
to 550 K. This has both pros and cons on sampling efficiency,
i.e., a favorable effect of allowing more active rearrangement of
monomers, and an unfavorable effect of inducing
aggregation-dispersion transition. While the latter
was not very ``sharp'' for the present system and thus well tractable,
it may pose a challenge for larger systems. Several approaches
have been proposed to deal with such cooperative transition
in REMD,\cite{REMD_coop}
which may also prove beneficial for the present purpose.

Although all-atom self-assembly simulations are limited
in both time and length scales, they are useful
for providing atomistic insights into key interactions
between monomers, which in turn serve as the basic information
for building (or refining) coarse-grained models.
We expect that REST will be useful particularly for such calculations.
Applying REST to larger supramolecular
systems remains a challenge for future study.

%------------------------------------------------------------------------
% @SI / @supp

\section*{SUPPLEMENTARY MATERIAL}
More computational details (with the definition of $\Nc$ and $\Lp$);
evolution of $\Nc$ for cMD at various temperatures;
potential energy distribution for each replica in T-REMD and REST;
additional statistical data and MD snapshots
for assembly structures; replica traversal in temperature space for T-REMD;
an electronic archive for the configuration
and topology files of the present BTA system.

%------------------------------------------------------------------------
% @acknowledgments

\vspace{-0.5cm}
\begin{acknowledgments}
The authors acknowledge support from JSPS Grant-in-Aid for Scientific Research
on Innovative Areas ``Dynamical ordering of biomolecular systems
for creation of integrated functions'' (Grant No. 25102002).
\end{acknowledgments}

%------------------------------------------------------------------------
% @ref / @bib

\bibliography{manuscript}

\end{document}